\documentclass{webofc}
\usepackage[varg]{txfonts}   

\usepackage{amsmath,graphicx,multirow,tabularx,physics}
\usepackage{davitex,slashed,listings}

\newcommand{\hp}{\hat{p}}
\newcommand{\hn}{\hat{n}}

\newcommand{\hs}{\hat{s}}
\newcommand{\hmu}{\muh}

\begin{document}

\title{QCD material parameters at zero and non-zero chemical potential from the lattice}

\author{\firstname{David Anthony} \lastname{Clarke}\inst{1}\fnsep\thanks{\email{clarke.davida@gmail.com}} \and
        \firstname{Jishnu} \lastname{Goswami}\inst{2}\and
        \firstname{Frithjof} \lastname{Karsch}\inst{3}\and
        \firstname{Peter} \lastname{Petreczky}\inst{4}
}

\institute{Department of Physics and Astronomy, University of Utah, Salt Lake City, Utah 84112, United States 
\and
           RIKEN Center for Computational Science, Kobe 650-0047, Japan 
\and
           Fakult\"at f\"ur Physik, Universit\"at Bielefeld, D-33615 Bielefeld, Germany
\and
Physics Department, Brookhaven National Laboratory, Upton, New York 11973, USA
          }

\abstract{%
Using an eighth-order Taylor expansion in baryon chemical potential, we recently 
obtained the (2+1)-flavor QCD equation of state (EoS) at non-zero conserved charge chemical 
potentials from the lattice. We focused on strangeness-neutral, isospin-symmetric QCD matter, 
which closely resembles the situation encountered in heavy-ion collision experiments. 
Using this EoS, we present here results on various QCD material parameters; 
in particular we compute the specific heat, speed of 
sound, and compressibility along appropriate lines of constant physics.
We show that in the entire range relevant for the beam energy scan at RHIC, 
the specific heat, speed of sound, and compressibility show no indication for 
an approach to critical behavior that one would expect close to a possibly existing 
critical endpoint.
}

\maketitle

\section{Introduction}\label{sec:intro}

A major goal of the experimental program on heavy-ion collisions (HIC) is to investigate 
transport and thermodynamic properties of strongly-interacting matter
in the plane of temperature $T$ and baryon chemical potential $\mu_B$.
Included among these properties are material parameters like the speed of sound,
the compressibility, and the specific heat.
The isentropic speed of sound $c_s^2$ is interesting, e.g., in the context of neutron stars
since the relationship between the star masses and radii is influenced by how $c_s^2$ 
changes with baryon number density $n_B$~\cite{Ozel:2016oaf}.
The isothermal speed of sound $c_T^2$ is also interesting for HIC,
as a new method to estimate $c_T^2$ in HIC 
has been recently suggested in Ref.~\cite{Sorensen:2021zme}.
Finally the isovolumetric specific heat $C_V$ can be related to the temperature 
fluctuations in HIC~\cite{Shuryak:1997yj}.

It is of special interest to probe
the $\mu_B$-$T$ plane for $\mu_B>0$, where a hypothesized first-order line separating
a hadronic gas phase and a quark-gluon plasma phase terminates in a critical
endpoint (CEP). Material parameters provide useful information about the nature
of the CEP. For example $c_s^2$ would drop to zero at
a true phase transition, and at a second-order transition, $C_V$ would show
a singularity.

Some of these material parameters have been previously calculated on the 
lattice at $\mu_B=0$ \cite{Gavai:2004se,Borsanyi:2013bia,HotQCD:2014kol}.
Here we present our ongoing calculations of these quantities at nonzero $\mu_B$.

\section{Strategy of the calculation}

We define
    $\hat{X}\equiv XT^{-k}$
with $k\in\Z$ chosen so that $\hat{X}$ is unitless. We expand the pressure $\hp$
in terms of the conserved charge chemical potentials $\hmu_B$, $\hmu_Q$, $\hmu_S$ as
\begin{equation}
\hp = \frac{1}{VT^3}\log\ZQCD(T,V,\hmu_B,\hmu_Q,\hmu_S) 
= \sum_{i,j,k=0}^\infty
\frac{\chi_{ijk}^{BQS}}{i!j!k!} \hmu_B^i \hmu_Q^j \hmu_S^k,
\label{eq:Pdefinition}
\end{equation}
where $\ZQCD$ is the QCD grand partition function, $V$ is the spatial volume, and
\begin{equation}
\chi_{ijk}^{BQS}\equiv \chi_{ijk}^{BQS}(T) =\left. 
\frac{\partial \hp}{\partial\hmu_B^i \partial\hmu_Q^j \partial\hmu_S^k}\right|_{\vec{\mu}=0} \; .
\label{eq:suscept}
\end{equation}
Other observables such as the entropy density $\hs$ and net-charge densities $\hn$
are derived from $\hp$ using standard thermodynamic relations.
To limit our analysis to the $\mu_B$-$T$ plane while focusing on the relevant 
physics of HIC, we impose constraints
    $n_S =0$  and $n_Q/n_B=r$.
In the following, partial derivatives are understood to be evaluated
at fixed $r$ and $n_S$. We focus on
\begin{equation}
    c_s^2=\atFixed{\pdv{p}{\epsilon}}{s/n_B},~~~~~
    c_T^2=\atFixed{\pdv{p}{\epsilon}}{T},~~~~~
    \kappa_s = \frac{1}{n_B} \atFixed{\pdv{n_B}{p}}{s/n_B},~~~~~
    C_V=T\atFixed{\pdv{s}{T}}{n_B},
\end{equation}
where $\epsilon$ is the energy density.
In our previous work~\cite{Bollweg:2022fqq} we computed $c_s^2$ using lattice data for various
$s/n_B$. To do this, we exploited the fact that
\begin{equation}
    c_s^2
    =\atFixed{\frac{\partial p/\partial T}{\partial \epsilon/\partial T}}{s/n_B}.
\end{equation}
We then interpolated $p$ and $\epsilon$ results simulated at various $T$ and computed
the derivative numerically. While this approach is straightforward, it is not ideal
to use interpolations since the numerical derivatives, especially higher-order ones,
are quite sensitive to the interpolation result. Since we estimate errors using a bootstrap
procedure, this can lead to substantially different estimates for the derivatives in each
bin and hence an artificially large error bar. 

Now we address these large statistical uncertainties by utilizing analytic formulas for the
material parameters in terms of cumulants, reducing the need to interpolate as much as possible.
Besides yielding more controlled uncertainties in the lattice data, we found this approach
increases numerical stability for our fixed $s/n_B$ HRG results, allowing us to extend our
calculations as low as $s/n_B=10$~\cite{Clarke:2022pfz}. When possible, our analytic
formulas are cross-checked against known thermodynamic relations; for instance
we find our expressions for $c_s$ and $\kappa_s$ to formally satisify
$
    \kappa_s^{-1}=c_s^2(\epsilon+p-\mu_Qn_Q-\mu_Sn_S).
$

\section{Computational setup}\label{sec:setup}

We use high-statistics data sets for
$(2+1)$-flavor QCD with degenerate light quark masses
$m_u=m_d\equiv m_l$ and a strange quark mass $m_s$
tuned so that $m_s/m_l=27$. These are the same
data sets as in Ref.~\cite{Bazavov:2017dus,Bollweg:2022rps}. 
We employed a HISQ action generated using 
$\simulat$~\cite{Bollweg:2021cvl,HotQCD:2023ghu}.
Temperatures above 180~MeV use data~\cite{Bazavov:2017dus}
with slightly heavier\footnote{This is known to have a negligible effect 
on the results~\cite{Bazavov:2011nk}.} light quarks, $m_s/m_l=20$.
In all cases results have been obtained on
lattices with aspect ratio $N_\sigma/N_\tau=4$.
In these proceedings, we present calculations only for the isospin-symmetric
case $r=0.5$. While $r=0.4$ is a more physically accurate choice,
choosing $r=0.5$ has the advantage of forcing $\muh_Q=0$, simplifying
some of the formulas. Moreover the quantitative differences between $r=0.4$
and $r=0.5$ EoS are generally mild~\cite{Goswami:2022nuu} and are hence
expected to have little impact on these parameters\footnote{Indeed, this
is what we found for the sound speeds~\cite{Clarke:2022pfz}.}.
We are often interested in the behavior of observables near the pseudocritical temperature
$\Tpc$. When indicated on figures, we take $\Tpc =156.5(1.5)$~MeV from Ref.~\cite{HotQCD:2018pds}.
Lines of constant $s/n_B$ and $n_B/n_0$, with nuclear matter density
$n_0=0.16/{\rm fm}^{3}$, are taken from Ref.~\cite{Bollweg:2022fqq}.
The \texttt{AnalysisToolbox}~\cite{toolbox} is used for HRG calculations,
spline fits, and bootstrapping. For the HRG model, we use the
QMHRG2020 list of hadron resonances \cite{Bollweg:2021vqf}.

\section{Results}\label{sec:results}

\begin{figure*}
\centering
\includegraphics[width=0.52\linewidth]{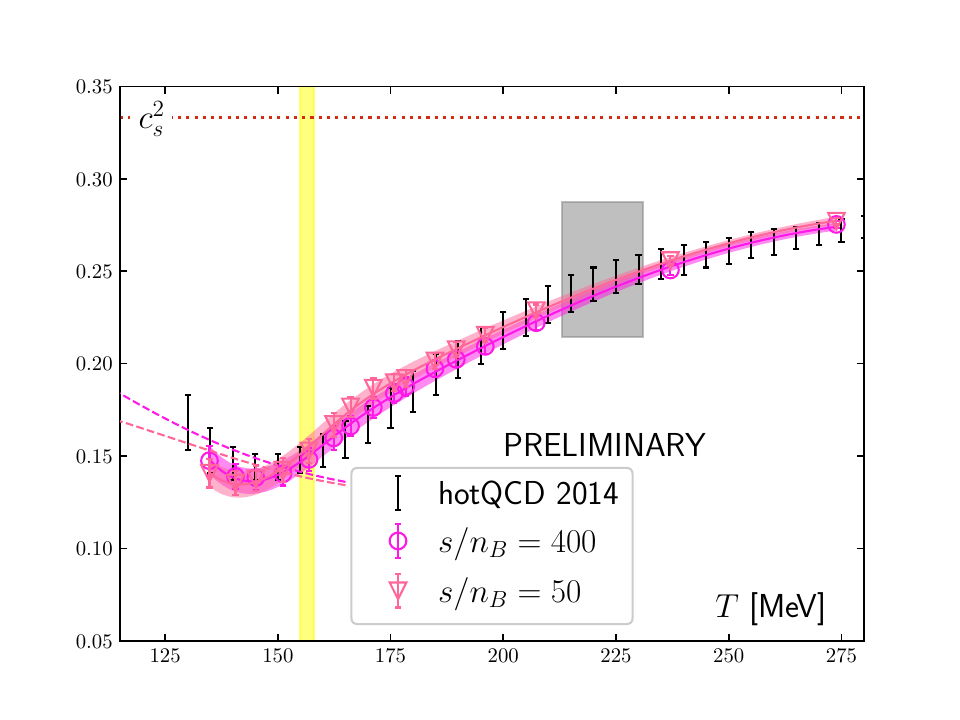}\hspace{-7mm}
\includegraphics[width=0.52\linewidth]{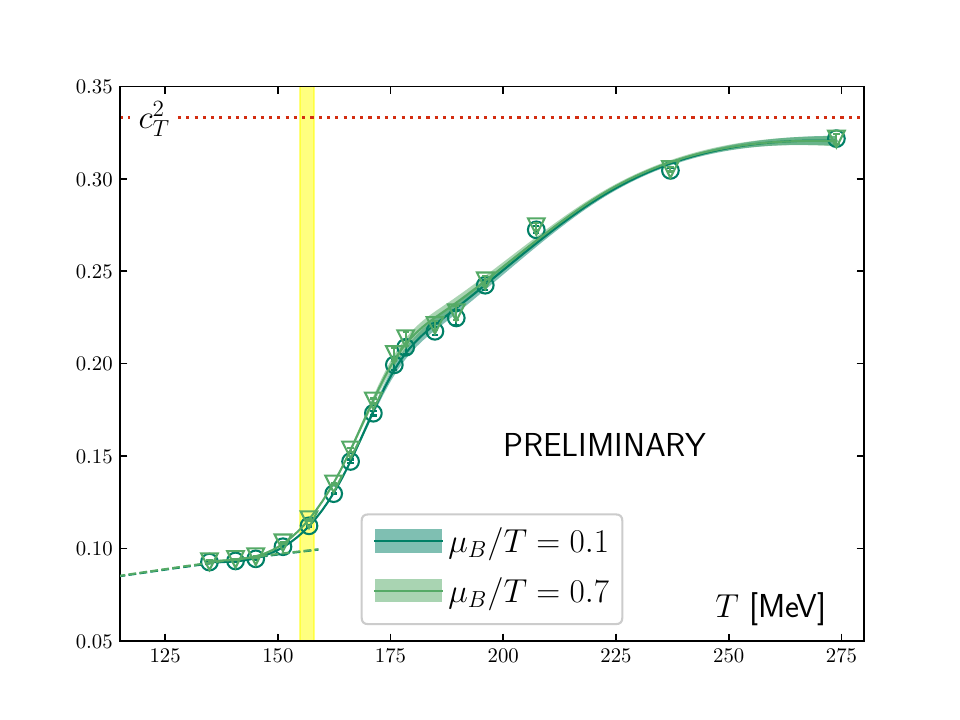}\\[-3mm]
\includegraphics[width=0.52\linewidth]{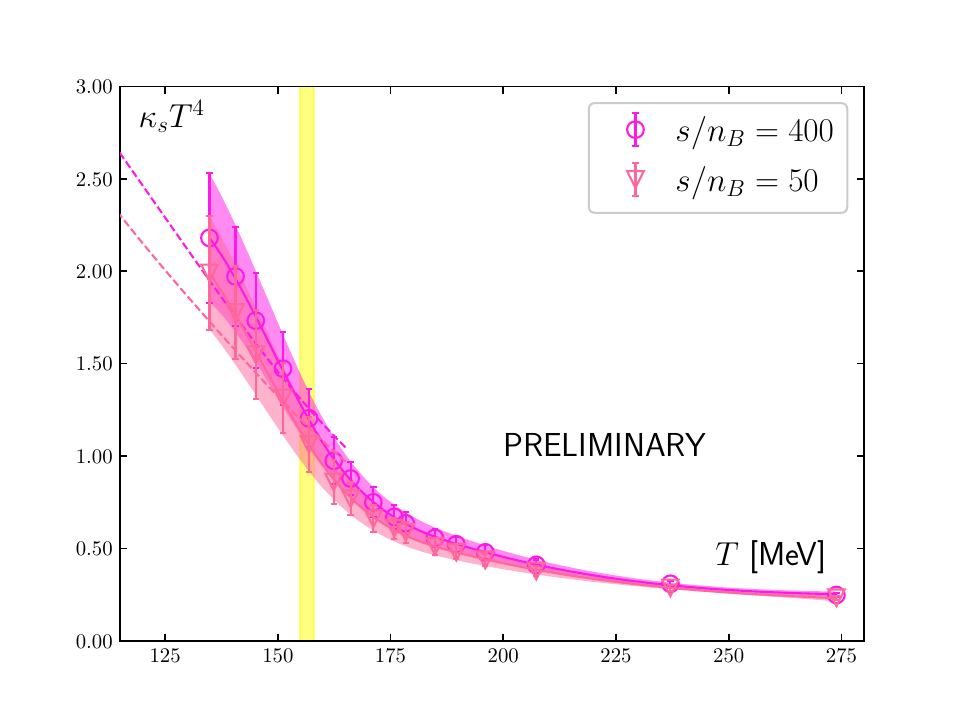}\hspace{-7mm}
\includegraphics[width=0.52\linewidth]{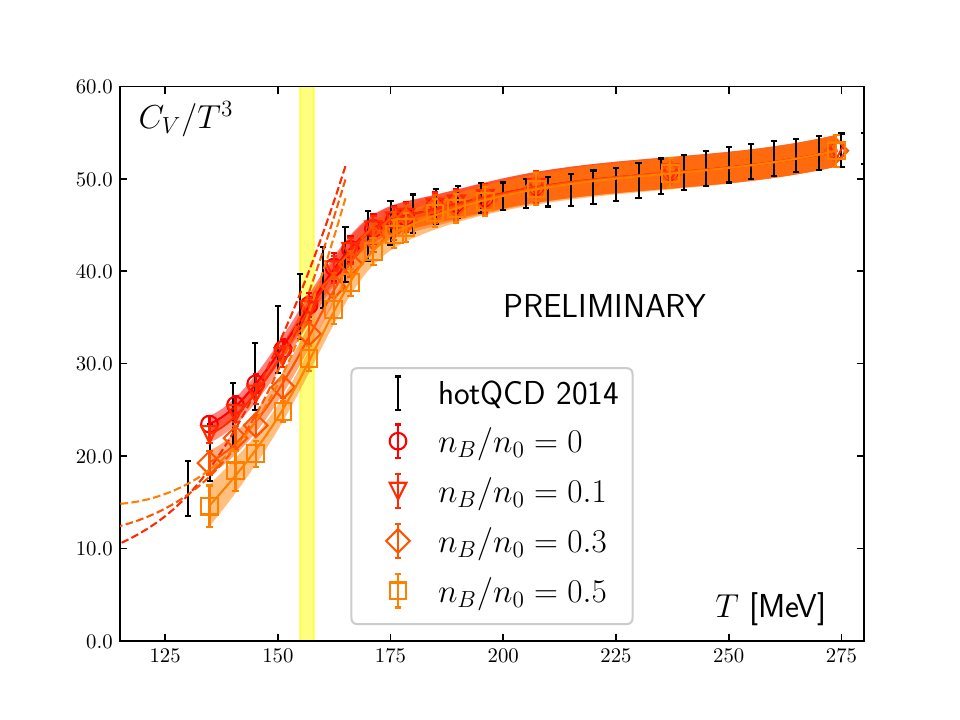}
\caption{Material parameters at $r=0.5$.
The bands are a spline
fit to the data and errors. The yellow band shows $\Tpc$.
The dashed lines show the HRG result. For the sound speeds,
the red, dotted line indicates the conformal limit. 
{\it Top left}: Isentropic speed of sound calculated using eqs.~(C3), (C4), and (C5) of 
Ref.~\cite{Bollweg:2022fqq}. The grey box indicates $c_s^2$ extracted from
hydrodynamic simulations using experimental data~\cite{Gardim:2019xjs}.
{\it Top right}: Isothermal speed of sound.
{\it Bottom left}: Isentropic compressibility.
{\it Bottom right}: Isovolumetric heat capacity. The hotQCD data, taken
from Ref.~\cite{HotQCD:2014kol}, were computed at $\mu_B=0$.}
\label{fig:materialParams}
\end{figure*}

In \figref{fig:materialParams} we show preliminary results for $c_s^2$, $c_T^2$, $\kappa_s$,
and $C_V$. Not all uncertainty has been included, hence error bands are 
mildly underestimated. Starting with isentropic observables, we note that because 
$n_B$ leads at $\order{\mu_B}$ while $s$ leads
with a constant, the limit $\mu_B\to\infty$ corresponds to $s/n_B\to0$.
Hence the left-hand plots show an at most mild dependence on $\mu_B$ in
the surveyed range. We see no indication of $c_s^2$ going to zero
within this range and hence no critical signature.
As has been seen already with $\mu_B=0$ calculations, 
$c_s^2$ overlaps with the estimate from Ref.~\cite{Gardim:2019xjs},
and both isentropic observables agree with our previous computation 
in Ref.~\cite{Bollweg:2022fqq}.
Turning to $c_T^2$, our preliminary results are similar to the preliminary
results of Ref.~\cite{Clarke:2022pfz}. Finally our results for $C_V$ at
$\mu_B=0$ agree with previous HotQCD results~\cite{HotQCD:2014kol}.
For all observables we see good agreement between lattice data and
HRG below $\Tpc$.

\section{Summary and outlook}\label{sec:outlook}

We presented the status of our ongoing calculation of various QCD material parameters.
The computation of the thermal expansion coefficient and
isobaric heat capacity are in progress, which besides being interesting
in their own right, will enable a few more
analytic cross-checks between the parameters.
For all our projects involving the QCD EoS, it will be useful to eventually
have continuum extrapolations for $6\nth$- and $8\nth$-order cumulants, but this
is a much more long-term goal.\\[1mm]

{\it Acknowledgements}--DAC was supported by the National Science Foundation under Grants PHY20-13064 and PHY23-10571.

\bibliography{bibliography.bib}

\end{document}